\documentclass[preprint,12pt,showpacs,showkeys,secnumarabic,aps]{revtex4}%
\usepackage{amsfonts}
\usepackage{amsmath}
\usepackage{amssymb}
\usepackage{graphicx}%
\begin{document}
\title{A lattice gas of prime numbers and the Riemann Hypothesis}
\author{Fernando Vericat}
\affiliation{Grupo de Aplicaciones Matem\'{a}ticas y Estad\'{\i}sticas de la Facultad de
Ingenier\'{\i}a (GAMEFI). Universidad Nacional de La Plata, Argentina}
\altaffiliation{Also at Instituto de F\'{\i}sica de L\'{\i}quidos y Sistemas Biol\'{o}gicos
(IFLYSIB)-CONICET-CCT La Plata, Argentina.}

\altaffiliation{E-mail: vericat@iflysib.unlp.edu.ar}

\begin{abstract}
In recent years, there has been some interest in applying ideas and methods
taken from Physics in order to approach several challenging mathematical
problems, particularly the Riemann Hypothesis. Most of these kind of
contributions are suggested by some quantum statistical physics problems or by
questions originated in chaos theory. In this article we show that the real
part of the non-trivial zeros of the Riemann zeta function extremizes the
grand potential corresponding to a simple model of one-dimensional classical
lattice gas, the critical point being located at 1/2 as the Riemann Hypothesis claims.

\end{abstract}

\maketitle

\section{Introduction}

The relation between physics and number theory has a long and fruitful history
as the two books referenced \cite{Luck1} and \cite{Waldschmidt1} suggest. This
link is in both directions: from number theory to physics and from physics to
number theory. At the beginning, the first direction was the more frequently
transited and so many number theoretic functions (and concepts in general)
were used in field theory, in classical and quantum statistical physics and
also in chaos theory. In this sense the Riemann zeta function\cite{Edwards1}
defined by $\zeta\left(  s\right)  :=\sum\nolimits_{n=1}^{\infty}n^{-s}$ in
the half-plane $\Re\left(  s\right)  >1$ and by analytical continuation in the
whole complex plane, plays an important role.

Given the ubiquitousness of $\zeta\left(  s\right)  $ in physics as well as in
mathematics itself, its behavior, particularly the properties of its zeros,
has been much studied. Nowadays we know that $\zeta\left(  s\right)  =0$ has
as trivial solutions the negative integers $-2k$ with $k\in$ $%
\mathbb{N}
$. The so-called non-trivial zeros are complex numbers most of whose
properties are also known. In an application of $\zeta\left(  s\right)  $ to a
number theoretic problem, specifically the distribution of the prime numbers,
Riemann conjectured more than 150 years ago that all the complex zeros have
1/2 as real part\cite{Edwards1}. This statement is the famous Riemann
Hypothesis whose proof has unsuccessfully involved many mathematicians since
then. This apparent inaccessibility to their proof from a full rigorous
mathematical point of view has, perhaps, motived the interest in applying
ideas and methods taken from physics in order to approach the Riemann
Hypothesis in particular as well as several other number theoretic problems.
In this way we can mention the work done around the idea of Hilbert and Polya,
proposed at the beginning of the past century, that the complex zeros of
$\zeta\left(  s\right)  $ constitute the spectrum of an operator
$\mathcal{R}=1/2\mathcal{I}+i\mathcal{H}$ where the Hamiltonian $\mathcal{H}$
is self-adjoint. Also, it is worth mentioning diverse applications of
statistical physics such as the use of random walks in order to approach the
Riemann Hypothesis\cite{Shlesinger1},\cite{Wolf1} or the search for
statistical mechanics models whose partition functions be related to
$\zeta\left(  s\right)  $ by using spin chains\cite{Knauf1} or gases of
harmonic oscillators\cite{Julia1}. These (and several others) contributions of
physics to number theory can be seen in the excellent review by Schumayer and
Hutchinson\cite{Schumayer1}.

In this article, following the way from physics to number theory, we give
support to the Riemann Hypothesis by extremizing the grand potential of a
simple classical one-dimensional lattice gas of prime numbers. In next section
we discuss the model and the equilibrium statistical mechanics functions that
we need for our analysis, particularly the grand potential of the system.
Section 3 is devoted to present the explicit formula given by Riemann for the
prime-counting function in terms of the zeros of the zeta function and, as a
natural consequence of the enunciation of the properties of these ones, to
state again the Riemann Hypothesis. Finally in section 4 we give support to
the Riemann Hypothesis from our model by showing that the real part of the
complex zeros of the zeta function\ must extremize the grand potential
obtained in section 2. Three appendices are considered in order to justify
some equations of the main text.

\section{Lattice gas of prime numbers}

Physically our model is very simple: a one-dimensional lattice gas in the
grand canonical ensemble. The lattice is an interval of the natural numbers:
$\left[  1,M\right]  \subset%
\mathbb{N}
$ where $M$ is large enough (eventually, in the thermodynamic limit,
$M\rightarrow\infty$). The system is in contact with a particles reservoir
characterized by the chemical potential $\mu$ and a heat reservoir at
temperature $T$. For $N$ particles the configuration of the system is given by
$\omega\equiv\left(  \omega_{N},N\right)  $ where $\omega_{N}\equiv\left(
i_{1},i_{2},\cdots,i_{N}\right)  $. The coordinates $i_{\alpha}$ $\left(
\alpha=1,2,\cdots,N\right)  $ take values in $\left[  1,M\right]  $ and $N$
ranges, in principle, between $0$ and $\infty$. The set of such
configurations, the configuration space, will be denoted $\Lambda$.

We assume that each site in the lattice can have at most one particle, so we
are considering a hard point pair potential%

\begin{equation}
u_{2}\left(  i_{\alpha},i_{\beta}\right)  =%
\genfrac{\{}{.}{0pt}{}{\infty\text{ \ \ \ if }i_{\alpha}=i_{\beta}}{0\text{
\ \ \ \ if }i_{\alpha}\neq i_{\beta}.}%
\text{ } \tag{1}\label{1}%
\end{equation}
Also we assume that the particles are subject to a one-point potential such
that they can just occupy sites in the lattice which are prime numbers:\vspace
{0in}%

\begin{equation}
u_{1}\left(  i_{\alpha}\right)  =%
\genfrac{\{}{.}{0pt}{}{0\text{ if }i_{\alpha}\text{ is prime \ \ \ \ \ \ }%
}{\infty\text{ if }i_{\alpha}\text{ is composite.}}
\tag{2}\label{2}%
\end{equation}
By the way, an example of explicit function with this form is\cite{Willans1}:
\[
u_{1}\left(  i_{\alpha}\right)  =-\log\left[  \sin^{2}\left\{  \pi\left[
\left(  i_{\alpha}-1\right)  !\right]  ^{2}/i_{\alpha}\right\}  /\sin
^{2}\left\{  \pi/i_{\alpha}\right\}  \right]  .
\]

The energy for the configuration $\omega$ is $u\left(  \omega\right)
=\sum_{\alpha=1}^{N}u_{1}\left(  i_{\alpha}\right)  +\sum_{\alpha<\beta}%
^{N}u_{2}\left(  i_{\alpha},i_{\beta}\right)  $.

Mathematically, a state of the system is a probability vector $\nu=\left(
\nu\left(  \omega\right)  \left\vert \omega\in\Lambda\right.  \right)  $. The
set of all states is denoted by $\mathcal{M}$. In the state $\nu$ the system
has the mean energy $\nu\left(  u\right)  :=\sum_{\omega\in\Lambda}\nu\left(
\omega\right)  u\left(  \omega\right)  $. The (grand) partition function of
$u$ is defined $\Xi\left(  \beta,\mu\right)  :=$ $\sum_{\omega\in\Lambda}%
\exp\left[  -\beta\left(  u\left(  \omega\right)  -\mu N\right)  \right]  $
where $\beta=\left(  k_{B}T\right)  ^{-1}$ with $k_{B}$ the Boltzmann
constant. For the parameters $\beta,M$ and $\mu$ the Gibbs measure is defined%

\begin{equation}
\nu_{0}\left(  \omega\right)  :=\frac{1}{\Xi\left(  \beta,M,\mu\right)  }%
\exp\left[  -\beta\left(  u\left(  \omega\right)  -\mu N\right)  \right]  .
\tag{3}\label{3}%
\end{equation}

Given a measure $\nu$ we have the entropy $H\left(  \nu\right)  :=-\sum
_{\omega\in\Lambda}\nu\left(  \omega\right)  \log\nu\left(  \omega\right)  $
and the grand potential%

\begin{equation}
\Omega\left[  \nu\right]  :=\nu\left(  u\right)  -\frac{1}{\beta}H\left(
\nu\right)  -\mu\nu\left(  N\right)  \tag{4}\label{4}%
\end{equation}
with $\nu\left(  N\right)  :=\sum_{\omega\in\Lambda}\nu\left(  \omega\right)
N$.

To introduce the notion of equilibrium state we consider a variational
principle for finite systems ($\left\vert \Lambda\right\vert $=finite)
according to which, for given energy $u$ and parameters $\beta,M$ and $\mu$,
the Gibbs measure satisfies%

\begin{equation}
\Omega\left[  \nu_{0}\right]  =-\frac{1}{\beta}\log\Xi\left(  \beta
,M,\mu\right)  =\inf_{\nu\in\mathcal{M}}\left(  \nu\left(  u\right)  -\frac
{1}{\beta}H\left(  \nu\right)  -\mu\nu\left(  N\right)  \right)  .
\tag{5}\label{5}%
\end{equation}
A measure that attains this infimum is called an equilibrium state. Gibbs
measure $\nu_{0}$ is thus an equilibrium state. The result given by
Eq.(\ref{5}) is easily demonstrated\cite{Georgii1} by using Jensen inequality
applied to the concave function $x\mapsto\ln x$ (see Appendix A). The
principle can be expressed saying that for any measure $\nu$ is $\Omega\left[
\nu\right]  \geqslant\Omega\left[  \nu_{0}\right]  =-\frac{1}{\beta}\log
\Xi\left(  \beta,M,\mu\right)  $ with equality if and only if $\nu=\nu_{0}$.

Turning to our model, the grand potential for the Gibbs state is easily
calculated if we explicitly write the grand partition function (
$z=\exp\left[  \beta\mu\right]  $):%

\[
\Xi\left(  \beta,M,\mu\right)  =%
{\displaystyle\sum\limits_{N=0}^{\infty}}
\frac{z^{N}}{N!}%
{\displaystyle\sum\limits_{i_{1}=1}^{M}}
{\displaystyle\sum\limits_{i_{2}=1}^{M}}
\cdots%
{\displaystyle\sum\limits_{i_{N}=1}^{M}}
\exp\left[  -\beta\left(  \sum_{\alpha=1}^{N}u_{1}\left(  i_{\alpha}\right)
+\sum_{\alpha<\beta}^{N}u_{2}\left(  i_{\alpha},i_{\beta}\right)  \right)
\right]
\]
and observe that, because of the limitation of occupation to just the prime
numbers (Eq.\ref{2}), the coordinates $i_{\alpha}$ $\left(  \alpha
=1,2,\cdots,N\right)  $ can take values only among the $\pi\left(  M\right)  $
prime numbers that exist in the interval $\left[  1,M\right]  $. Also, because
of the impenetrability of the particles (Eq.\ref{1}), the choice can be made
of $\pi\left(  M\right)  !/\left[  \pi\left(  M\right)  -N\right]  !$ manners.
We obtain%

\begin{equation}
\Xi\left(  \beta,M,\mu\right)  =%
{\displaystyle\sum\limits_{N=0}^{\pi\left(  M\right)  }}
\binom{\pi\left(  M\right)  }{N}z^{N}=\left(  1+z\right)  ^{\pi\left(
M\right)  } \tag{6}\label{6}%
\end{equation}
and%

\begin{equation}
\Omega\left[  \nu_{0}\right]  =-\frac{1}{\beta}\log\left(  1+z\right)
^{\pi\left(  M\right)  }. \tag{7}\label{7}%
\end{equation}

\section{Riemann Hypothesis}

The prime-counting function $\pi\left(  x\right)  $ is the number of prime
numbers less or equal than $x$. In his 1859 classic memoir to the Berlin
Academy of Sciences (see for example the monograph by Edwards\cite{Edwards1}
for a translation), Riemann gave an explicit formula for $\pi\left(  x\right)
$ in terms of the zeros and the pole of the analytical continuation of
$\zeta\left(  s\right)  $\cite{Edwards1}:%

\begin{equation}
\pi\left(  x\right)  =\overline{\pi}\left(  x\right)  +\widetilde{\pi}\left(
x\right)  , \tag{8}\label{8}%
\end{equation}
where the smooth part, to which contribute the pole at $s=1$ and the integer
zeros of $\zeta\left(  s\right)  $, is given by%

\begin{equation}
\overline{\pi}\left(  x\right)  =\sum_{n=1}^{\infty}\frac{\mu\left(  n\right)
}{n}\left[  \text{Li}\left(  x^{1/n}\right)  -\sum_{k=1}^{\infty}%
\text{Li}\left(  x^{-2k/n}\right)  \right]  , \tag{9}\label{9}%
\end{equation}
with Li$\left(  x\right)  $ the logarithmic integral function and $\mu\left(
n\right)  $ the M\"{o}bius function defined%

\[
\text{ \ \ \ \ \ \ \ \ \ \ \ \ \ \ \ \ \ \ \ \ }\mu\left(  n\right)  =\left\{
\begin{array}
[c]{c}%
0\text{ if }n\text{ has one or more repeated prime
factors\ \ \ \ \ \ \ \ \ \ \ \ \ \ \ \ \ }\\
\left(  -1\right)  ^{r}\text{ if }n\text{ is a product of }r\text{ distinct
primes \ \ \ \ \ \ \ \ \ \ \ \ \ \ \ \ \ \ \ \ \ }\\
\text{ \ \ }1\text{ if }n=1.\text{
\ \ \ \ \ \ \ \ \ \ \ \ \ \ \ \ \ \ \ \ \ \ \ \ \ \ \ \ \ \ \ \ \ \ \ \ \ \ \ \ \ \ \ \ \ \ \ \ \ \ \ \ \ \ \ \ \ \ \ \ \ \ \ \ \ \ }%
\end{array}
\right.
\]
The complex zeros, on the other hand, contribute to the oscillatory part

\begin{equation}
\widetilde{\pi}\left(  x\right)  =-2\text{Re}\sum_{n=1}^{\infty}\frac
{\mu\left(  n\right)  }{n}\sum_{\alpha=1}^{\infty}\text{Li}\left(  x^{\left(
\sigma_{\alpha}+it_{\alpha}\right)  /n}\right)  , \tag{10}\label{10}%
\end{equation}
where the real numbers $\sigma_{\alpha}$ and $t_{\alpha}$ are the real and
imaginary part, respectively, of the complex number $\rho_{\alpha}%
=\sigma_{\alpha}+it_{\alpha}$\ that verifies $\zeta\left(  \rho_{\alpha
}\right)  =0.$

For $x$ large enough the non-oscillatory and oscillatory parts can be written,
respectively, as%
\begin{equation}
\overline{\pi}\left(  x\right)  \approx\frac{x}{\log x} \tag{11}\label{11}%
\end{equation}
and (see Appendix B)%

\begin{equation}
\widetilde{\pi}\left(  x\right)  \approx-\frac{2}{\log x}%
{\displaystyle\sum\limits_{\alpha=1}^{\infty}}
\frac{x^{\sigma_{\alpha}}}{\left(  \sigma_{\alpha}^{2}+t_{\alpha}^{2}\right)
}\left[  \sigma_{\alpha}\cos\left(  t_{\alpha}\log x\right)  +t_{\alpha}%
\sin\left(  t_{\alpha}\log x\right)  \right]  . \tag{12}\label{12}%
\end{equation}
For simplicity, from now on we will consider this last situation; so $M$ will
be assumed large enough.

As we mentioned in the Introduction, the properties of the complex or non
trivial zeros of the Riemann%
\'{}%
s zeta function has been extensively studied\cite{Edwards1}. For example it
has been demonstrated that there are infinitely many complex zeros and that
all of them lie inside the region \ $0<\Re\left(  s\right)  <1$ and are
symmetric about the real axis $\Im\left(  s\right)  =0$. Also is well known
that if $\xi\left(  s\right)  :=$ $\frac{s}{2}\left(  s-1\right)  \pi^{-s/2}$
$\mathit{\Gamma}\left(  \frac{s}{2}\right)  \zeta(s)$ then the non trivial
zeros of $\zeta(s)$ are precisely the zeros of $\xi\left(  s\right)  $ and
since $\xi\left(  s\right)  =\xi\left(  1-s\right)  $, we have that the
complex zeros of $\zeta(s)$ are symmetric with respect to the so called
critical line $\Re\left(  s\right)  =1/2$. The Riemann Hypothesis is the
statement that all the complex zeros have their real part exactly on the
critical line: $\sigma_{\alpha}=1/2$ $\forall\alpha$. Although billions of
complex zeros have been numerically calculated\cite{Odlyzko1} confirming all
of them this conjecture, the full demonstration of the general validity of the
Riemann Hypothesis remains still open.

\section{Supporting the Riemann Hypothesis from the model}

We wish to analyze the behavior of the equilibrium grand partition potential
(Eq.\ref{7}), with $\pi\left(  M\right)  $ given by Eqs.(\ref{8}-\ref{12}), as
a function of the quantities $\sigma_{\alpha}\in\left(  0,1\right)  $
\ $\left(  \alpha=1,2,\cdots\right)  $. These numbers must be thought as
possible values for the real part of the zeta function zeros and our goal
would be to identify the true ones.

We focus into a generic zero $\rho_{\gamma}$ and his complex conjugate
$\rho_{\gamma}^{\ast}$ (which is known is also a zero) and observe that, since
Riemann actually didn't know the reliable values of the complex zeros, Eqs.
(\ref{10}) and (\ref{12}) must be taken as general enough as to contemplate
the possibility that, associated to $t_{\gamma}=\Im\left(  \rho_{\gamma
}\right)  $, both cases, $\sigma_{\gamma}$ $\neq$ $1-\sigma_{\gamma}$ and
$\sigma_{\gamma}$ $=$ $1-\sigma_{\gamma}$, can be considered for $\Re\left(
\rho_{\gamma}\right)  $.

Explicitly we rewrite Eq.(\ref{7}) for $\beta,M$ and $\mu$ fixed%

\begin{equation}
\Omega\left(  \beta,M,\mu;\sigma_{\gamma}\right)  =-\frac{1}{\beta}\log\left(
1+z\right)  ^{\pi\left(  M;\sigma_{\gamma}\right)  }, \tag{13}\label{13}%
\end{equation}
with%

\begin{equation}
\pi\left(  M;\sigma_{\gamma}\right)  =\overline{\pi}\left(  M\right)
+\widetilde{\pi}_{\neq\gamma}\left(  M\right)  +\widetilde{\pi}_{\gamma
}\left(  M;\sigma_{\gamma}\right)  . \tag{14}\label{14}%
\end{equation}
Here the term $\widetilde{\pi}_{\neq\gamma}\left(  M\right)  $ includes all
the complex zeros except those labelled $\gamma$ and we assume that its
contribution to the prime-counting function is the true one. We write the
remaining term:%

\begin{equation}
\widetilde{\pi}_{\gamma}\left(  M;\sigma_{\gamma}\right)  \approx\left\{
\begin{array}
[c]{c}%
-\frac{4}{\log M}\left(  \frac{M^{\sigma_{\gamma}}}{\sigma_{\gamma}%
^{2}+t_{\gamma}^{2}}\left[  \sigma_{\gamma}\cos\left(  t_{\gamma}\log
M\right)  +t_{\gamma}\sin\left(  t_{\gamma}\log M\right)  \right]  +\right.
\text{
\ \ \ \ \ \ \ \ \ \ \ \ \ \ \ \ \ \ \ \ \ \ \ \ \ \ \ \ \ \ \ \ \ \ \ \ \ \ \ \ \ \ \ \ \ \ \ \ \ \ \ \ \ \ \ \ \ \ \ \ \ \ \ \ \ \ \ \ \ \ \ \ \ \ \ \ \ \ \ \ \ \ \ \ \ \ \ \ \ \ \ \ \ \ \ \ \ \ \ \ \ \ \ \ \ \ \ \ \ \ \ \ \ \ \ \ \ \ \ \ \ \ \ \ \ \ \ \ \ \ \ \ \ \ \ \ }%
\\
\text{\ }\\
\left.  \frac{M^{\left(  1-\sigma_{\gamma}\right)  }}{\left(  1-\sigma
_{\gamma}\right)  ^{2}+t_{\gamma}^{2}}\left[  \left(  1-\sigma_{\gamma
}\right)  \cos\left(  t_{\gamma}\log M\right)  +t_{\gamma}\sin\left(
t_{\gamma}\log M\right)  \right]  \right)  \text{\ \ for }\sigma_{\gamma}%
\neq1-\sigma_{\gamma}%
\text{\ \ \ \ \ \ \ \ \ \ \ \ \ \ \ \ \ \ \ \ \ \ \ \ \ \ \ \ \ \ \ \ \ \ \ \ \ \ \ \ \ \ \ \ \ \ \ \ \ \ \ \ \ \ \ \ \ \ \ \ \ \ \ \ \ \ \ \ \ \ \ \ \ \ \ \ \ \ \ \ \ \ \ \ \ \ \ \ \ \ \ \ \ \ \ \ \ \ \ \ \ \ \ \ \ \ \ \ \ \ \ \ \ \ \ \ \ }%
\\%
\begin{array}
[c]{c}%
\\
-\frac{4}{\log M}\frac{M^{1/2}}{\frac{_{1}}{4}+t_{\gamma}^{2}}\left[  \frac
{1}{2}\cos\left(  t_{\gamma}\log M\right)  +t_{\gamma}\sin\left(  t_{\gamma
}\log M\right)  \right]  \text{ \ \ \ \ \ \ \ for\ }\sigma_{\gamma}%
=1-\sigma_{\gamma},
\end{array}
\text{
\ \ \ \ \ \ \ \ \ \ \ \ \ \ \ \ \ \ \ \ \ \ \ \ \ \ \ \ \ \ \ \ \ \ \ \ \ \ \ \ \ \ \ \ \ \ \ \ \ \ \ \ \ \ \ \ \ \ \ \ \ \ \ \ \ \ \ \ \ \ \ \ \ \ \ \ \ \ \ \ \ \ \ \ \ \ \ \ \ \ \ \ \ \ \ \ \ \ \ \ \ \ \ \ \ \ \ \ \ \ \ \ \ }%
\end{array}
\right.  \tag{15}\label{15}%
\end{equation}
where we have taken into account the fact that if $\zeta\left(  \sigma
_{\gamma}+it_{\gamma}\right)  =0$ then also $\zeta\left(  1-\sigma_{\gamma
}+it_{\gamma}\right)  =0$, $\zeta\left(  \sigma_{\gamma}-it_{\gamma}\right)
=0$ and\ $\zeta\left(  1-\sigma_{\gamma}-it_{\gamma}\right)  =0$.

Special cases can make the analysis even simpler. For $M=\left[  M_{1}\right]
$ ($\left[  \bullet\right]  $ integer part of $\bullet$) with $M_{1}%
=\exp[\frac{\pi}{2}\left(  4n-1\right)  /\left\vert t_{\gamma}\right\vert ]$
and $M=\left[  M_{2}\right]  $ with $M_{2}=\exp[\frac{\pi}{2}\left(
4n-3\right)  /\left\vert t_{\gamma}\right\vert ]$ ($n\in%
\mathbb{N}
$ large enough) and taking into account that $\left\vert t_{\alpha}\right\vert
>14.1347$ for all the non trivial zeros of the zeta function\cite{Edwards1},
we have%

\begin{equation}
\widetilde{\pi}_{\gamma}\left(  M;\sigma_{\gamma}\right)  \approx\left\{
\begin{array}
[c]{c}%
\pm\frac{4}{\left\vert t_{\gamma}\right\vert \log M_{1,2}}\left(
M_{1,2}^{\sigma_{\gamma}}+M_{1,2}^{\left(  1-\sigma_{\gamma}\right)  }\right)
\text{ for }\sigma_{\gamma}\neq1-\sigma_{\gamma}\\
\pm\frac{4}{\left\vert t_{\gamma}\right\vert \log M_{1,2}}M_{1,2}^{1/2}\text{
\ \ \ \ \ \ \ \ \ \ \ \ \ \ \ \ \ \ \ \ \ for }\sigma_{\gamma}=1-\sigma
_{\gamma},
\end{array}
\right.  \tag{16}\label{16}%
\end{equation}
where the upper sign corresponds to $M_{1}$. Note the avoidable discontinuity
in the function $\widetilde{\pi}_{\gamma}\left(  M;\sigma_{\gamma}\right)  $
when $\sigma_{\gamma}=1-\sigma_{\gamma}$.

Using Eq.(\ref{13}) together with Eq.(\ref{15}) -or the particular cases given
by Eq.(\ref{16})- is easy to see that the grand potential has a unique
extremum (minimum or maximum depending on $M$) at the interval $0<\sigma
_{\gamma}<1$. From a heuristic point of view it is reasonable to expect that
this extremum (maximum = unstable equilibrium; minimum = stable equilibrium)
be reached when $\sigma_{\gamma}$ takes the value $\left(  \sigma_{\gamma
}\right)  _{true}$ that gives, through Eq.(\ref{14}), the correct
time-invariant number of prime numbers lying inside the interval $\left[
1,M\right]  $.

The same conclusion can be achieved more formally by considering the problem
from a dynamic point of view. To this, we firstly take into account that
fixing a given value for $\sigma_{\gamma}$ can be thought as imposing a
constraint that limits (through $\pi\left(  M;\sigma_{\gamma}\right)  $) the
number of the accessible states to the system. If this constraint is removed,
so $\sigma_{\gamma}$ is leaved to freely vary in the interval $\left(
0,1\right)  $, then the equilibrium probability $P(\sigma_{\gamma})$ that the
system be in states with the parameter taking values in the interval between
$\sigma_{\gamma}$ and $\sigma_{\gamma}+\delta\sigma_{\gamma}$ behaves as (see
Appendix C)%

\begin{equation}
P(\sigma_{\gamma})\propto%
\genfrac{\{}{.}{0pt}{}{\exp\left[  -\beta\Omega\left(  \beta,M,\mu
;\sigma_{\gamma}\right)  \right]  \text{ \ if \ }\sigma_{\gamma}\in\left(
0,1\right)  }{\text{ \ \ \ \ }0\text{
\ \ \ \ \ \ \ \ \ \ \ \ \ \ \ \ \ \ \ \ \ \ \ \ \ \ \ if \ }\sigma_{\gamma
}\notin\left(  0,1\right)  .}
\tag{17}\label{17}%
\end{equation}
We can then arbitrarily introduce a second random variable such that
(\ref{17}) be the marginal of the joint probability of both ones. We use as
new variable a Maxwellian one $\overset{\cdot}{\sigma_{\gamma}}$ so that the
bi-dimensional random variable $\left(  \sigma_{\gamma},\overset{\cdot}%
{\sigma_{\gamma}}\right)  $ describes the phase space of an hypothetical
particle of mass $m$ moving in a force field with potential function given by%
\begin{equation}
U\left(  \sigma_{\gamma}\right)  =%
\genfrac{\{}{.}{0pt}{}{\Omega\left(  \beta,M,\mu;\sigma_{\gamma}\right)
\text{ \ if \ }\sigma_{\gamma}\in\left(  0,1\right)  }{\text{ \ \ \ }%
\infty\text{ \ \ \ \ \ \ \ \ \ \ \ \ \ \ \ \ if \ }\sigma_{\gamma}%
\notin\left(  0,1\right)  .}
\tag{18}\label{18}%
\end{equation}

In this picture, $P(\sigma_{\gamma})$ can be thought as the marginal of the
joint probability distribution%

\begin{equation}
P(\sigma_{\gamma},\overset{\cdot}{\sigma_{\gamma}})=\frac{1}{Z\left(
\beta\right)  }\exp\left\{  -\beta\left[  U\left(  \sigma_{\gamma}\right)
+\frac{1}{2}m\overset{\cdot}{\sigma_{\gamma}}^{2}\right]  \right\}  .
\tag{19}\label{19}%
\end{equation}
Here the normalization constant $Z\left(  \beta\right)  $ is the single
particle canonical partition function%

\begin{equation}
Z\left(  \beta\right)  =\iint e^{-\beta\mathcal{H}\left(  \sigma_{\gamma
},\overset{\cdot}{\sigma_{\gamma}}\right)  }d\sigma_{\gamma}d\overset{\cdot
}{\sigma_{\gamma}}, \tag{20}\label{20}%
\end{equation}
where $\mathcal{H}\left(  \sigma_{\gamma},\overset{\cdot}{\sigma_{\gamma}%
}\right)  $ denotes the Hamiltonian%

\begin{equation}
\mathcal{H}\left(  \sigma_{\gamma},\overset{\cdot}{\sigma_{\gamma}}\right)
=\frac{1}{2}m\overset{\cdot}{\sigma_{\gamma}}^{2}+U\left(  \sigma_{\gamma
}\right)  . \tag{21}\label{21}%
\end{equation}
The Hamilton equations of motion for the variables $\sigma_{\gamma}$ and
$\overset{\cdot}{\sigma_{\gamma}}$ are%

\begin{equation}
\overset{\cdot}{\sigma_{\gamma}}=\frac{\partial\mathcal{H}}{m\partial
\overset{\cdot}{\sigma_{\gamma}}},\text{ \ \ \ \ \ \ \ }m\overset{\cdot\cdot
}{\sigma_{\gamma}}=-\frac{\partial\mathcal{H}}{\partial\sigma_{\gamma}%
},\text{\ \ \ \ \ \ \ } \tag{22}\label{22}%
\end{equation}
which yield the Newton equation%

\begin{equation}
m\frac{\overset{\cdot}{d\sigma_{\gamma}}}{dt}=-\frac{dU\left(  \sigma_{\gamma
}\right)  }{d\sigma_{\gamma}}. \tag{23}\label{23}%
\end{equation}

If we assume that initially the particle is placed at the position
$\sigma_{\gamma_{0}}=$ $\left(  \sigma_{\gamma}\right)  _{extr}$ that
extremizes $\Omega\left(  \beta,M,\mu;\sigma_{\gamma}\right)  $ with velocity
$\overset{\cdot}{\sigma_{\gamma}}_{0}=0$, then the equation of motion has as
the unique obvious solution the isolated fixed point $\sigma_{\gamma}\left(
t\right)  =\left(  \sigma_{\gamma}\right)  _{extr}\forall t$. Taking into
account that for any other pair of initial conditions the solution is
time-dependent and that in our lattice gas model the number $\pi\left(
M;\sigma_{\gamma}\right)  $ of prime numbers less or equal than $M$ does not
change with time but remains constant, we deduce that it must be $\left(
\sigma_{\gamma}\right)  _{true}\equiv\left(  \sigma_{\gamma}\right)  _{extr}$.
One then infers that $\left(  \sigma_{\gamma}\right)  _{true}$ should verify%

\begin{equation}
\left.  \frac{\partial\Omega\left(  \beta,M,\mu;\sigma_{\gamma}\right)
}{\partial\sigma_{\gamma}}\right\vert _{\sigma_{\gamma}=\left(  \sigma
_{\gamma}\right)  _{true}}=-\frac{1}{\beta}\log\left(  1+z\right)  \left.
\frac{\partial\widetilde{\pi}_{\gamma}\left(  M;\sigma_{\gamma}\right)
}{\partial\sigma_{\gamma}}\right\vert _{\sigma_{\gamma}=\left(  \sigma
_{\gamma}\right)  _{true}}=0. \tag{24}\label{24}%
\end{equation}
Derivation of expressions (\ref{15}) or (\ref{16}) shows that this equation
has the form $\left[  f\left(  \sigma_{\gamma}\right)  -f\left(
1-\sigma_{\gamma}\right)  \right]  _{\sigma_{\gamma}=\left(  \sigma_{\gamma
}\right)  _{true}}=0$ (with $f\left(  \sigma_{\gamma}\right)  \neq
0$;\ $f\left(  1-\sigma_{\gamma}\right)  \neq0$) which is clearly verified by
$\left(  \sigma_{\gamma}\right)  _{true}=1-\left(  \sigma_{\gamma}\right)
_{true}=1/2$. Note that the maximum (minimum) achieved at 1/2 when the
discontinuity is avoided is smaller (larger) than the corresponding values
taken by the function $\widetilde{\pi}_{\gamma}\left(  M;\sigma_{\gamma
}\right)  $ in (\ref{15}) or (\ref{16}) when $\sigma_{\gamma}=1-\sigma
_{\gamma}$.

Because the zero $\rho_{\gamma}$ that we have considered is a generic one, we
would conclude that $\left(  \sigma_{\alpha}\right)  _{true}=1/2$
$\forall\alpha$, say that, in fact, the real part of all non-trivial zeros of
Riemann's zeta function lie on the critical line.\bigskip

\textbf{Acknowledgements}

I thank very much Dr. Osvaldo H. Scalise for very useful discussions during
the preparation of the manuscript. Support of this work by Consejo Nacional de
Investigaciones Cient\'{\i}ficas y T\'{e}cnicas, Universidad Nacional de La
Plata and Agencia Nacional de Promoci\'{o}n Cient\'{\i}fica y Tecnol\'{o}gica
of Argentina is greatly appreciated. The author is a member of CONICET.

\bigskip\newpage

\begin{center}
APPENDIX A: Variational principle in the grand canonical ensemble
\end{center}

Here we follow Georgii \cite{Georgii1} to proof \ the result given by
Eq.(\ref{5}). For any arbitrary measure $\nu\in\mathcal{M}$, we have, using
Jensen inequality together with the definition given by Eq.(\ref{4}):%

\begin{align}
\Omega\left[  \nu\right]   &  =-\frac{1}{\beta}\sum_{\omega\in\Lambda}%
\nu\left(  \omega\right)  \left[  -\log\nu\left(  \omega\right)  -\beta
u\left(  \omega\right)  +\beta\mu N\right] \nonumber\\
&  =-\frac{1}{\beta}\sum_{\omega\in\Lambda}\nu\left(  \omega\right)  \log
\frac{e^{-\beta\left[  u\left(  \omega\right)  -\mu N\right]  }}{\nu\left(
\omega\right)  }\nonumber\\
&  \geq-\frac{1}{\beta}\log\sum_{\omega\in\Lambda}\nu\left(  \omega\right)
\frac{e^{-\beta\left[  u\left(  \omega\right)  -\mu N\right]  }}{\nu\left(
\omega\right)  }, \tag{A1}\label{25}%
\end{align}
or, taking into account Eq. (\ref{3}),%

\begin{equation}
\Omega\left[  \nu\right]  \geq-\frac{1}{\beta}\log\Xi=\Omega\left[  \nu
_{0}\right]  . \tag{A2}\label{26}%
\end{equation}
The equality holds if and only if the random variable \ $\omega\mapsto
\frac{e^{-\beta\left[  u\left(  \omega\right)  -\mu N\right]  }}{\nu\left(
\omega\right)  }$ is constant, i.e., if $\nu=\nu_{0}$ since, in this case,
according to Eq.(\ref{3}), is $\frac{e^{-\beta\left[  u\left(  \omega\right)
-\mu N\right]  }}{\nu\left(  \omega\right)  }=\Xi=$ $constant$.

\begin{center}
\bigskip

APPENDIX B: Equation (\ref{12})
\end{center}

For $x$ large enough, is\cite{Edwards1} Li$\left(  x^{\rho_{\alpha}}\right)
\approx x^{\rho_{\alpha}}/\log x^{\rho_{\alpha}}$ and the dominant term in
Eq.(\ref{10}) is that corresponding to $n=1$:%

\begin{align}
\widetilde{\pi}\left(  x\right)   &  \approx-2\text{Re}\sum_{\alpha=1}%
^{\infty}\text{Li}\left(  x^{\sigma_{\alpha}+it_{\alpha}}\right) \nonumber\\
&  \approx-2\text{Re}\sum_{\alpha=1}^{\infty}\frac{x^{\sigma_{\alpha
}+it_{\alpha}}}{\log x^{\sigma_{\alpha}+it_{\alpha}}}. \tag{B1}\label{27}%
\end{align}
Using $x^{it_{\alpha}}=e^{it_{\alpha}\log x}$ we have%

\begin{align}
\widetilde{\pi}\left(  x\right)   &  \approx-\frac{2}{\log x}\text{Re}%
\sum_{\alpha=1}^{\infty}\frac{x^{\sigma_{\alpha}}\left[  \cos\left(
t_{\alpha}\log x\right)  +i\sin\left(  t_{\alpha}\log x\right)  \right]
}{\left(  \sigma_{\alpha}+it_{\alpha}\right)  }\nonumber\\
&  \approx-\frac{2}{\log x}\text{Re}\sum_{\alpha=1}^{\infty}\frac
{x^{\sigma_{\alpha}}\left[  \cos\left(  t_{\alpha}\log x\right)  +i\sin\left(
t_{\alpha}\log x\right)  \right]  \left(  \sigma_{\alpha}-it_{\alpha}\right)
}{\sigma_{\alpha}^{2}+t_{\alpha}^{2}}, \tag{B2}\label{28}%
\end{align}
from where Eq.(\ref{12}) follows.

\begin{center}
\bigskip

\bigskip

APPENDIX C: The Probability $P\left(  \sigma_{\gamma}\right)  $
\end{center}

In this appendix we deduce Eq. (\ref{17}) for $P\left(  \sigma_{\gamma
}\right)  $ by applying to our case the method of Reif\cite{Reif1}. We call
$A_{t}=A\cup A_{r}$ the total isolated system made up of the lattice gas and
the reservoir at temperature $T_{r}$ and chemical potential $\mu_{r}$. In
equilibrium the temperature and chemical potential\ of $A$ equals those of
$A_{r}$: $T=T_{r}$; $\mu=\mu_{r}$. We are assuming that $A$ (and consequently
$A_{t}$) depends on $\sigma_{\gamma}$ through $\pi\left(  M;\sigma_{\gamma
}\right)  $ so that if $\sigma_{\gamma}$ changes from a given value
$\sigma_{\gamma}=\sigma_{\gamma1}$ to a generic one $\sigma_{\gamma}$, then
the thermodynamical functions of $A$ ( i.e. $H$, $\nu\left(  u\right)  $,
$\nu\left(  N\right)  $ or $\Omega$) will also change (for example $\Delta
H=H\left(  \sigma_{\gamma}\right)  -$ $H\left(  \sigma_{\gamma1}\right)  $)
and so do the corresponding thermodynamical functions of $A_{t}$ (e.g. $\Delta
H_{t}=H_{t}\left(  \sigma_{\gamma}\right)  -$ $H_{t}\left(  \sigma_{\gamma
1}\right)  $).

Because of the additivity of the entropy, we can write its variation in the
total system as $\Delta H_{t}=\Delta H+\Delta H_{r}$. But the variation in the
reservoir entropy is given by $\Delta H_{r}=-\beta Q$ where $Q$ denotes the
heat absorbed by $A$ from $A_{r}$ and $\beta=\left(  k_{B}T\right)
^{-1}=\left(  k_{B}T_{r}\right)  ^{-1}$. According to the first law of
thermodynamics applied to $A$, the heat $Q$ can be written%

\begin{equation}
Q=\Delta\nu\left(  u\right)  -\mu\Delta\nu\left(  N\right)  +W \tag{C1}%
\label{29}%
\end{equation}
where: $\Delta\nu\left(  u\right)  $ is the variation in the internal energy
of $A$, $\Delta\nu\left(  N\right)  $ the variation in the mean particles
number of $A$ and $W$ the\ \ work done by $A$ along the process.

Then we have%

\begin{align*}
\Delta H_{t}  &  =\Delta H-\beta Q=\beta\left[  \frac{\Delta H}{\beta}-\left(
\Delta\nu\left(  u\right)  -\mu\Delta\nu\left(  N\right)  +W\right)  \right]
=\\
&  \text{\ \ \ \ \ \ \ \ \ \ \ \ \ \ \ \ }-\beta\left[  \Delta\left(
\nu\left(  u\right)  -\frac{H}{\beta}-\mu\nu\left(  N\right)  \right)
+W\right]  ,
\end{align*}
or, remembering the definition given by Eq. (\ref{4}) in the text:%

\begin{equation}
\Delta H_{t}=-\beta\left[  \Delta\Omega+W\right]  . \tag{C2}\label{30}%
\end{equation}

Assuming that the external parameters of $A$ are fixed we have $W=0$, so that
the total entropy will change%

\begin{equation}
\Delta H_{t}=H_{t}\left(  \sigma_{\gamma}\right)  -H_{t}\left(  \sigma
_{\gamma1}\right)  =-\beta\left[  \Omega\left(  \sigma_{\gamma}\right)
-\Omega\left(  \sigma_{\gamma1}\right)  \right]  . \tag{C3}\label{31}%
\end{equation}

Now, taking into account that the total system is isolated so that all the
micro-states (configurations $\omega$) which are accessible to $A_{t}$ have
the same probability, we have\cite{Reif1}%

\begin{equation}
P\left(  \sigma_{\gamma}\right)  \propto\mathcal{N}_{t}\left(  \sigma_{\gamma
}\right)  =e^{H_{t}\left(  \sigma_{\gamma}\right)  }, \tag{C4}\label{32}%
\end{equation}
where $\mathcal{N}_{t}\left(  \sigma_{\gamma}\right)  $ denotes the number of
micro-states accessible to $A_{t}$ when the parameter takes values between
$\sigma_{\gamma}$ and $\sigma_{\gamma}+\delta\sigma_{\gamma}$. It is worth
mentioning that if the equilibrium state of $A_{t}$ when the parameter is
fixed at $\sigma_{\gamma}$ is $\nu_{t}\left(  \sigma_{\gamma}\right)  =\left(
\nu_{t}\left(  \omega;\sigma_{\gamma}\right)  \left\vert \omega\in\Lambda
_{t}\left(  \sigma_{\gamma}\right)  \right.  \right)  $ with $\nu_{t}\left(
\omega;\sigma_{\gamma}\right)  =1/\mathcal{N}_{t}\left(  \sigma_{\gamma
}\right)  $ where $\mathcal{N}_{t}\left(  \sigma_{\gamma}\right)  =\left\vert
\Lambda_{t}\left(  \sigma_{\gamma}\right)  \right\vert $ and $\Lambda
_{t}\left(  \sigma_{\gamma}\right)  $ the configuration space, then the
entropy is $H_{t}\left(  \sigma_{\gamma}\right)  =-\sum_{\omega\in\Lambda
_{t}\left(  \sigma_{\gamma}\right)  }\nu_{t}\left(  \omega;\sigma_{\gamma
}\right)  \log\nu_{t}\left(  \omega;\sigma_{\gamma}\right)  =\log
\mathcal{N}_{t}\left(  \sigma_{\gamma}\right)  $ from where the equality of
Eq.(\ref{32}) follows. Replacing $H_{t}\left(  \sigma_{\gamma}\right)  $ in
Eq.(\ref{32}) by the expression obtained from Eq.(\ref{31}) and including into
the proportionality constant all the terms which are functions of the
arbitrary value $\sigma_{\gamma1}$ we have $P\left(  \sigma_{\gamma}\right)
\propto e^{-\beta\Omega\left(  \sigma_{\gamma}\right)  \text{ \ }}$, say
Eq.(\ref{17}) in the text.

\end{document}